\documentclass[10pt,letterpaper]{article}
\usepackage[top=0.85in,left=2.75in,footskip=0.75in]{geometry}

\usepackage{amsmath,amssymb}

\usepackage{changepage}

\usepackage{textcomp,marvosym}

\usepackage{cite}

\usepackage{nameref,hyperref}

\usepackage[right]{lineno}

\usepackage[nopatch=eqnum]{microtype}
\DisableLigatures[f]{encoding = *, family = * }

\usepackage[table]{xcolor}

\usepackage{array}

\newcolumntype{+}{!{\vrule width 2pt}}

\newlength\savedwidth

\raggedright
\usepackage[aboveskip=1pt,labelfont=bf,labelsep=period,justification=raggedright,singlelinecheck=off]{caption}

\makeatletter
\renewcommand{\@biblabel}[1]{\quad#1.}
\makeatother

\usepackage{lastpage,fancyhdr,graphicx}
\usepackage{epstopdf}
\fancyheadoffset[L]{2.25in}
\fancyfootoffset[L]{2.25in}
\usepackage{orcidlink}
\usepackage[skipempty]{credits}%
\usepackage{csquotes}
\usepackage{ulem}

\usepackage{glossaries}
\glsenableentrycount 
\let\gls\cgls 
\let\glspl\cglspl

\newacronym{nfdi}{NFDI}{National Research Data Infrastructure}
\newacronym{eosc}{EOSC}{European Open Science Cloud}
\newacronym{fair}{FAIR}{Findable, Accessible, Interoperable, Reusable}
\newacronym{gdpr}{GDPR}{General Data Protection Regulation}
\newacronym{w3c}{W3C}{World Wide Web Consortium}
\newacronym[longplural=Memoranda of Understanding]{mou}{MOU}{Memorandum of Understanding}
\newacronym{dua}{DUA}{Data Usage Agreement}
\newacronym{odrl}{ODRL}{Open Digital Rights Language}
\newacronym{etl}{ETL}{Extract, Transform, Load}
\newacronym{ffde}{FFDE}{FAIR and federated Data Ecosystems for interdisciplinary Research}

\begin{document}
\vspace*{0.2in}

\begin{flushleft}
{\Large
\textbf\newline{Towards FAIR and federated Data Ecosystems for interdisciplinary Research} 
}
\newline
%
Sebastian Beyvers%
\orcidlink{0000-0002-9747-7096}
\textsuperscript{1}, %
Jannis Hochmuth%
\orcidlink{0009-0004-4382-4760}
\textsuperscript{1}, %
Lukas Brehm%
\orcidlink{0009-0008-0887-0015}
\textsuperscript{1}, %
Maria Hansen%
\orcidlink{0009-0001-3265-8654}
\textsuperscript{1},
Alexander Goesmann%
\orcidlink{0000-0002-7086-2568}
\textsuperscript{1}, %
Frank F\"{o}rster%
\orcidlink{0000-0003-4166-5423}
\textsuperscript{1,2*}%
\\
\bigskip
\textbf{1} Bioinformatics and Systems Biology, Justus Liebig University Giessen, 35390 Giessen, Hesse, Germany
\\
\textbf{2} Bioinformatics Core Facility, Justus Liebig University Giessen, 35390 Giessen, Hesse, Germany
\\
\bigskip

%
%





* Frank.Foerster@cb.jlug.de

\end{flushleft}
\section*{Abstract}
Scientific data management is at a critical juncture, driven by exponential data growth, increasing cross-domain dependencies, and a severe reproducibility crisis in modern research.
Traditional centralized data management approaches are not only struggle with data volume, but also fail to address the fragmentation of research results across domains, hampering scientific reproducibility, and cross-domain collaboration, while raising concerns about data sovereignty and governance.
Here we propose a practical framework for FAIR and federated Data Ecosystems that combines decentralized, distributed systems with existing research infrastructure to enable seamless cross-domain collaboration.
Based on established patterns from data commons, data meshes, and data spaces, our approach introduces a layered architecture consisting of governance, data, service, and application layers.
Our framework preserves domain-specific expertise and control while facilitating data integration through standardized interfaces and semantic enrichment.
Key requirements include adaptive metadata management, simplified user interaction, robust security, and transparent data transactions.
Our architecture supports both compute-to-data as well as data-to-compute paradigms, implementing a decentralized peer-to-peer network that scales horizontally.
By providing both a technical architecture and a governance framework, FAIR and federated Data Ecosystems enables researchers to build on existing work while maintaining control over their data and computing resources, providing a practical path towards an integrated research infrastructure that respects both domain autonomy and interoperability requirements.


\section*{Author summary}

\nolinenumbers

\section{Introduction}\label{sec:introduction}

The exponential growth of scientific data, coupled with increasing cross-domain dependencies, has created an urgent need for innovative data management solutions \cite{demchenko2013}. While next-generation sequencing technologies revolutionize genomic research \cite{katz2021}, they also highlight a critical challenge: our current infrastructure struggles to handle not just the volume of data, but the complex web of relationships between research outputs across domains. Traditional solutions such as data warehouses, data lakes or even domain specific repositories, designed for centralized control, are becoming increasingly impractical \cite{nagesian2019}.

This challenge extends beyond mere storage and processing. The reproducibility crisis in modern research \cite{miyakawa2020no, ioannidis2005most}, exacerbated by fragmented data management practices, demands immediate action. While initiatives like the German \gls{nfdi}\cite{hartl2021nationale}, European \gls{eosc}\cite{ayris2016realising}, Gaia-X \cite{braud2021road} or 
data commons initiatives, like the Australian Research Data Commons \cite{barker2019australian} or the NIH Data Commons \cite{grossman2016} have begun addressing these challenges, a comprehensive solution remains elusive.

Cloud technologies offer promising capabilities for elastic resource allocation \cite{alzakholi2020}, enabling more efficient and sustainable research practices \cite{agrawal2020, patel2015}. However, relying solely on centralized cloud providers creates new challenges around data sovereignty and privacy. Recent initiatives such as data commons \cite{grossman2016} and data spaces \cite{otto2022evolution} demonstrate the potential of distributed approaches aligned with \gls{fair} principles \cite{wilkinson2016fair}. Yet these solutions often operate in isolation, creating new silos rather than true cross-domain integration.

We propose \gls{ffde}: a practical framework that combines the benefits of distributed systems with the reality of existing research infrastructure. This approach integrates established domain-specific repositories with support for intermediate research outputs—including experimental data, analysis workflows, and computational results. By providing both the technical architecture and governance framework for cross-domain collaboration, \gls{ffde} enable researchers to build upon existing work while maintaining control over their data and computational resources.

This paper outlines the key requirements for such a system and presents a concrete architectural approach to realize this vision.
We illustrate how combining proven patterns from data commons, data spaces, and cloud computing can create a foundation for future scientific collaboration—one that respects both the sovereignty of individual research groups and the need for seamless data integration across domains.

\section{Drawing from Existing Architectural Patterns}\label{subsec:architecturalpatterns}

Several architectural patterns have emerged to address different aspects of research data management. Among these, three patterns offer particularly relevant concepts for cross-domain collaboration. Data Commons \cite{grossman2016} demonstrate how shared infrastructure can collocate data with computational tools to enable reproducible workflows. The Data Mesh paradigm \cite{wider2023decentralized, goedegebuure2024data} illustrates the value of domain-oriented ownership, treating data as a product and enabling research group autonomy. Data spaces \cite{otto2022evolution} provide frameworks for secure data exchange through standardized interfaces.

\gls{ffde} draw from these and other patterns selectively: incorporating shared infrastructure concepts similar to Data Commons, domain-specific governance approaches inspired by Data Mesh, and standardized interfaces comparable to Data Spaces. This combination enables access to existing repositories, incorporation of new data sources, and cross-domain discovery while preserving specialized research practices. By synthesizing proven concepts, we aim to facilitate cross-domain scientific collaboration without compromising the autonomy of individual research communities.

\section{FAIR and federated Data Ecosystems}\label{sec:fairds}

Data spaces offer promising solutions for data sovereignty and governance, yet their adoption in scientific contexts remains limited. Current implementations demand significant technical expertise and typically focus on specific domains like health, energy, or automotive industries. This domain-specificity, while valuable for industry applications, creates barriers for broader scientific collaboration.

Modern research increasingly transcends traditional domain boundaries. For instance, findings from fluid dynamics can improve weather simulations, which in turn influence agricultural research on resilient plant breeds \cite{welsh2006post}. As we cannot predict which domains will need to collaborate in the future, research infrastructure must support flexible, cross-domain integration while preserving domain expertise and data sovereignty.%

We propose a \gls{fair} and federated Data Ecosystem approach as a foundation for the upcoming generation of research data infrastructure. By combining the governance strengths of Data spaces with practical accessibility, this approach demonstrates how established domain-specific repositories can be integrated with support for unpublished research outputs. The resulting infrastructure accommodates diverse data types, from raw experimental data to intermediate results, analysis procedures, and computational workflows, while preserving domain autonomy.

\subsection{Requirements}\label{subsec:requirements}

To enable a scalable, secure, and interoperable domain-agnostic data ecosystems, several key requirements must be addressed. These requirements balance technical capabilities with practical usability while ensuring broad adoption across research communities.

\paragraph{Governance}
Many domains already have dedicated governance structures that enforce unique rules and standards. The architecture need to accommodate this and allow multiple governance structures to coexist. In addition, a common set of baseline rules must be defined to which all entities must adhere. Clear transparency about the governance paradigms applied have to be communicated, both humanly understandable and machine-actionable. For privacy reasons, this information can be limited to those policies that directly affect public or external access.

\paragraph{Adaptability of Metadata Management}
The metadata management of data spaces should be adaptable to different scientific domains. While World Wide Web Consortium (W3C) \cite{w3c} standards and well-defined ontologies are ideal, practical constraints often require the separation of technical metadata from semantic metadata. This separation allows data owners full control over their semantics and allows any existing data source to participate. The required technical metadata contains only the core information about who wants to share what data and under what conditions, as well as basic information about how data records are related to each other and where the data is stored.

\paragraph{Simplifying User Interaction}
Widespread adoption is a primary goal to make the concept of domain-agnostic data ecosystems viable. To achieve this, user interaction must be as frictionless as possible, which can be addressed by facilitating the seamless integration of existing data repositories and by integrating widely accepted interfaces. Technical adapters play a crucial role by transforming various existing data exchange formats into a unified set of well-established and widely recognized standards to ensure ease of use and compliance with FAIR principles.

\paragraph{Data Sovereignty and Security Measures}
Maximizing data sovereignty means implementing robust technical systems to control access. Advanced security measures such as encryption and continuous monitoring are required to prevent unauthorized access and data breaches, as well as to comply with data protection regulations such as the European \gls{gdpr} \cite{thegdpr2018}.
All data protection systems should be multi-layered, and the final decision should always be made by data owners. Decentralization plays a critical role in empowering data owners to do this. Ideally, data transfer should take place on a peer-2-peer basis, without central authorities.

\paragraph{Transparency in Data Transactions}
Systems should provide data owners and consumers with clear visibility into data access, use, and compliance with privacy and ethical standards. Everyone should have a clear understanding of what is happening with their data. This requires a well-established authentication system to ensure that everyone knows who they are dealing with.

\subsection{Architectural Components}\label{subsec:architecturalcomponents}

The following sections outline the key components that form the backbone of our proposed architecture and specify some of the ideas that could be used to realize these components.

\begin{figure*}[ht]
  \includegraphics[width=\textwidth]{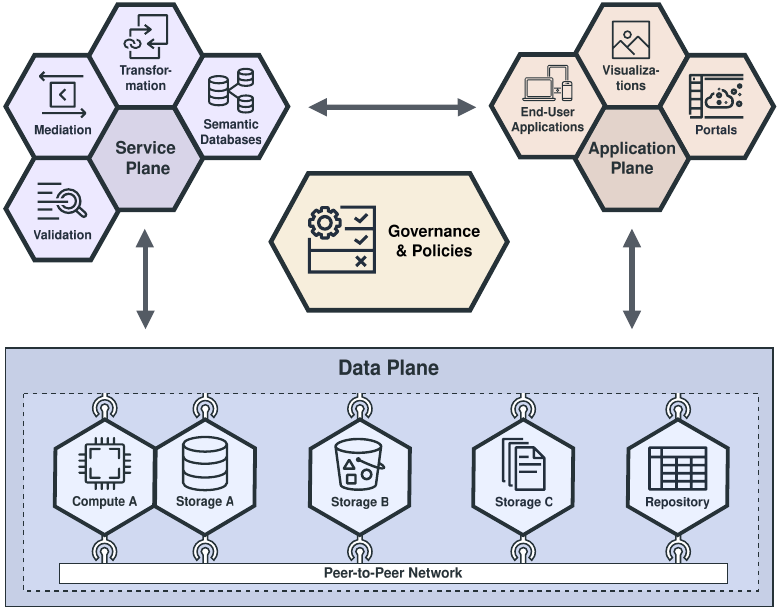}
  \caption{Schematic overview of the cloud-based service architecture, illustrating the interplay between data services, application services, governance structures and the underlying data plane. Storage and compute nodes are interconnected with each other through an internal peer-to-peer network and are orchestrated by the cloud management layer; connections are established by adapters.%
  \label{fig:layered-architecture}}
\end{figure*}

\paragraph{Governance and Access Control Plane}\mbox{} \\
Decentralized governance operates through distinct community clusters, representing research networks organized by institutional type, research domain, or geographic region. Each cluster implements standardized data sharing agreements that define common principles for data access, usage rights, and ethical compliance. These agreements establish baseline requirements while preserving flexibility for institutions to implement additional controls based on their specific contexts.

The access control mechanism operates through a chain of digital certificates and policy enforcement points. At the community level, a federated identity management system (such as eduGAIN \cite{lopez2006edugain} for academic institutions) validates researcher credentials and institutional affiliations. Data owners then apply granular access controls through attribute-based access control \cite{abac} policies, defining permissions based on research purpose and data sensitivity levels.

Inter-institutional trust relationships are formalized through cryptographically signed \glspl{mou}, supported by machine-readable \glspl{dua} in standardized formats like \gls{odrl} \cite{odrl}. This framework allows organizations to establish fast-track approval processes for trusted partners while maintaining comprehensive audit trails.

\paragraph{Data Plane}
The Data Plane implements a decentralized peer-to-peer network for distributed data management, going beyond existing technologies (like IPFS \cite{ipfs}) by adding rich metadata handling and advanced search capabilities essential for scientific data. It operates through two main components: Data Locations (sovereign storage units spanning multiple platforms) and Compute Nodes (processing resources that connect to existing data sources via standardized adapters). This creates a dynamically interlinked network of data sources that abstracts the underlying complexity from researchers, providing them with a seamless interface to distributed resources.

The architecture supports both compute-to-data and data-to-compute paradigms, optimizing distribution based on various requirements including processing capabilities, storage capacity, network bandwidth, privacy constraints, and geographical location. Nodes within the same domain or organization can establish tighter integration with shared vocabularies and optimized data transfer protocols, while maintaining standardized interfaces for cross-domain interactions. The system scales horizontally - adding new nodes linearly increases storage and processing capacity without central bottlenecks. While decentralized in nature, the infrastructure maintains governance through federated trust domains and standardized policies, allowing organizations to retain full control over their data assets while participating in the broader network.

The network automatically maps relationships between datasets and adapts its topology as nodes join or leave, with built-in location awareness enabling efficient resource allocation and data sovereignty compliance. This flexibility allows research infrastructure providers to easily integrate new storage and compute resources as their needs grow, without disrupting existing workflows or exposing the complexity to end users. The standardized interfaces ensure compatibility across different platforms while preserving each node's autonomy over its data and processing capabilities, creating a truly distributed research data infrastructure that balances openness with institutional control.

\paragraph{Data Service Plane}
The Data Service Plane transforms the distributed data sources of the data plane into an integrated collection of secondary data resources through automated \gls{etl} processes and semantic enrichment. These \gls{etl} pipelines continuously extract data from the distributed network, apply necessary transformations for standardization and quality control, and load the processed data into purpose-built secondary databases optimized for specific research queries and domains.

The plane leverages domain-specific ontologies and controlled vocabularies to harmonize data across sources. For example, in life sciences, terminologies like Gene Ontology \cite{go} and Disease Ontology \cite{do} are used to standardize metadata and establish semantic relationships between datasets. This semantic layer enables sophisticated query capabilities and automated inference of relationships between different data elements.

While maintaining the distributed Data Plane as the authoritative source of truth, the Data Service Plane creates and maintains specialized secondary databases optimized for common research access patterns. These derived databases significantly improve query performance and data accessibility while ensuring eventual consistency with primary sources through automated synchronization processes. For instance, aggregated genomic variant databases can be automatically updated as new sequencing data becomes available in the distributed network, maintaining a balance between data freshness and query efficiency.

\paragraph{Application Plane}
The Application Plane leverages both the Data and Service planes to construct sophisticated user-facing interfaces, transforming the distributed infrastructure into accessible research tools. It combines direct access to raw data sources from the Data Plane with optimized secondary databases from the Service Plane to deliver responsive and intuitive research applications. This dual approach enables both high-performance access to preprocessed data and direct exploration of primary sources when needed.

This plane implements a diverse set of interfaces to accommodate different user needs and technical expertise levels. These include web-based frontends with interactive data visualization dashboards, data portals, RESTful APIs for programmatic data access, command-line interfaces for batch processing, and plugin frameworks for integration with common tools. By abstracting the underlying distributed architecture behind these familiar interfaces, the plane significantly reduces the technical barrier to entry, allowing researchers to focus on their scientific questions rather than managing distributed systems, data formats, or compute resources, while giving them the option to reference a wide range of data sets from different fields.

\section{Conclusion}\label{sec:conclusion}

The architectural components and requirements outlined in this paper represent essential building blocks for future research data ecosystems. By combining distributed architectures with practical accessibility, we have described an approach that offers a viable path toward true cross-domain scientific collaboration. The proposed layered architecture—comprising governance, data, service, and application planes—provides the necessary flexibility to accommodate diverse research needs while maintaining data sovereignty and security.
The strength of this approach lies in its ability to bridge existing domain-specific repositories with emerging collaborative requirements. Rather than replacing established systems, it enhances them through standardized interfaces and semantic enrichment, enabling researchers to leverage both specialized tools and cross-domain capabilities.%

This acknowledgment highlights the growing importance of breaking down traditional domain boundaries to enable future scientific breakthroughs through data connectivity and analysis. However, realizing this potential depends on the active engagement of the research community. Research institutions should critically assess their existing data infrastructure in relation to the requirements outlined in this paper, identifying gaps and opportunities to strengthen cross-domain collaboration. Domain experts play a crucial role in shaping the foundations for meaningful data integration by contributing their knowledge to the development and refinement of domain-specific ontologies and metadata standards. At the same time, infrastructure providers must work towards implementing standardized interfaces that adhere to the architectural principles described, ensuring compatibility and seamless integration within the broader research ecosystem. Furthermore, funding bodies have a key responsibility in recognizing the value of cross-domain data sharing and interoperability, actively supporting initiatives that promote the long-term sustainability of an interconnected research landscape. Only through collective effort and commitment can these possibilities be fully realized, paving the way for innovative discoveries and a more integrated approach to scientific inquiry.%

The future of scientific research depends on our ability to break down data silos while respecting domain expertise and data sovereignty. The architectural components and requirements presented here provide guidance for achieving this balance, but success relies on widespread adoption and continuous refinement by the research community. The time to act is now—as the volume and complexity of scientific data continue to grow, the need for effective cross-domain collaboration becomes increasingly critical.

\section*{Funding}
SB and FF are funded by the German Network for Bioinformatics Infrastructure (de.NBI) (W-de.NBI-010).
JH and LB are funded by the German Research Foundation DFG under the grant agreement number 460129525 (NFDI4Microbiota). The project is part of NFDI, the National Research Data Infrastructure Program in Germany.
MH is funded by the German Research Foundation DFG under the grant agreement number 442032008 (NFDI4Biodiversity). The project is part of NFDI, the National Research Data Infrastructure Program in Germany.
Open Access funding enabled and organized by Projekt DEAL.

\section*{Author contributions}
\credit{SB}{1,0,1,0,1,1,0,1,1,0,1,1,1,1}
\credit{JH}{1,0,0,0,1,1,0,1,1,0,1,1,0,1}
\credit{LB}{1,0,0,0,1,1,0,1,1,0,1,1,0,1}
\credit{AG}{0,0,0,1,0,0,1,1,0,1,0,0,0,1}
\credit{FF}{1,0,0,0,1,1,1,1,1,1,1,1,1,1}
\insertcreditsstatement{}

\section*{Competing interests}
The authors declare that they have no competing interests.

\nolinenumbers


\begin{thebibliography}{10}

  \bibitem{demchenko2013}
  Demchenko Y, Grosso P, de~Laat C, Membrey P.
  \newblock Addressing big data issues in Scientific Data Infrastructure.
  \newblock In: 2013 International Conference on Collaboration Technologies and
    Systems (CTS); 2013. p. 48--55.
  
  \bibitem{katz2021}
  Katz K, Shutov O, Lapoint R, Kimelman M, Brister J, O’Sullivan C.
  \newblock {The Sequence Read Archive: a decade more of explosive growth}.
  \newblock Nucleic Acids Research. 2021;50(D1):D387--D390.
  \newblock doi:{10.1093/nar/gkab1053}.
  
  \bibitem{nagesian2019}
  Nargesian F, Zhu E, Miller RJ, Pu KQ, Arocena PC.
  \newblock Data lake management: challenges and opportunities.
  \newblock Proc VLDB Endow. 2019;12(12):1986–1989.
  \newblock doi:{10.14778/3352063.3352116}.
  
  \bibitem{miyakawa2020no}
  Miyakawa T.
  \newblock No raw data, no science: another possible source of the
    reproducibility crisis.
  \newblock Molecular brain. 2020;13:1--6.
  
  \bibitem{ioannidis2005most}
  Ioannidis JP.
  \newblock Why most published research findings are false.
  \newblock PLoS medicine. 2005;2(8):e124.
  
  \bibitem{hartl2021nationale}
  Hartl N, W{\"o}ssner E, Sure-Vetter Y.
  \newblock Nationale Forschungsdateninfrastruktur (NFDI).
  \newblock Informatik Spektrum. 2021;44(5):370--373.
  
  \bibitem{ayris2016realising}
  Commission E, for Research DG, Innovation.
  \newblock Realising the European open science cloud.
  \newblock Publications Office; 2016.
  
  \bibitem{braud2021road}
  Braud A, Fromentoux G, Radier B, Le~Grand O.
  \newblock The road to European digital sovereignty with Gaia-X and IDSA.
  \newblock IEEE network. 2021;35(2):4--5.
  
  \bibitem{barker2019australian}
  Barker M, Wilkinson R, Treloar A.
  \newblock The Australian research data commons.
  \newblock Data science journal. 2019;18:44--44.
  
  \bibitem{grossman2016}
  Grossman RL, Heath A, Murphy M, Patterson M, Wells W.
  \newblock A Case for Data Commons: Toward Data Science as a Service.
  \newblock Computing in Science \& Engineering. 2016;18(5):10--20.
  \newblock doi:{10.1109/MCSE.2016.92}.
  
  \bibitem{alzakholi2020}
  Alzakholi O, haji L, Shukur H, Zebari R, Abas S, Sadeeq M.
  \newblock Comparison Among Cloud Technologies and Cloud Performance.
  \newblock Journal of Applied Science and Technology Trends. 2020;1(2):40--47.
  \newblock doi:{10.38094/jastt1219}.
  
  \bibitem{agrawal2020}
  Agrawal MN, Saini MJK, Wankhede P.
  \newblock Review on green cloud computing: A step towards saving global
    environment.
  \newblock International Journal of Engineering Research \& Technology.
    2020;8(5):1--4.
  
  \bibitem{patel2015}
  Patel YS, Mehrotra N, Soner S.
  \newblock Green cloud computing: A review on Green IT areas for cloud computing
    environment.
  \newblock In: 2015 International Conference on Futuristic Trends on
    Computational Analysis and Knowledge Management (ABLAZE); 2015. p. 327--332.
  
  \bibitem{otto2022evolution}
  Otto B.
  \newblock The evolution of data spaces.
  \newblock In: Designing Data Spaces: The Ecosystem Approach to Competitive
    Advantage. Springer International Publishing Cham; 2022. p. 3--15.
  
  \bibitem{wilkinson2016fair}
  Wilkinson MD, Dumontier M, Aalbersberg IJ, Appleton G, Axton M, Baak A, et~al.
  \newblock The FAIR Guiding Principles for scientific data management and
    stewardship.
  \newblock Scientific data. 2016;3(1):1--9.
  \newblock doi:{https://doi.org/10.1038/sdata.2016.18}.
  
  \bibitem{wider2023decentralized}
  Wider A, Verma S, Akhtar A.
  \newblock Decentralized data governance as part of a data mesh platform:
    concepts and approaches.
  \newblock In: 2023 IEEE International Conference on Web Services (ICWS). IEEE;
    2023. p. 746--754.
  
  \bibitem{goedegebuure2024data}
  Goedegebuure A, Kumara I, Driessen S, Van Den~Heuvel WJ, Monsieur G, Tamburri
    DA, et~al.
  \newblock Data mesh: a systematic gray literature review.
  \newblock ACM Computing Surveys. 2024;57(1):1--36.
  
  \bibitem{welsh2006post}
  Welsh E, Jirotka M, Gavaghan D.
  \newblock Post-genomic science: cross-disciplinary and large-scale
    collaborative research and its organizational and technological challenges
    for the scientific research process.
  \newblock Philosophical Transactions of the Royal Society A: Mathematical,
    Physical and Engineering Sciences. 2006;364(1843):1533--1549.
  
  \bibitem{w3c}
  Consortium WWW. {World Wide Web Consortium (W3C)}; 2024.
  \newblock \url{https://www.w3.org/}.
  
  \bibitem{thegdpr2018}
  Commission E, for Justice DG, Consumers.
  \newblock The GDPR – New opportunities, new obligations – What every
    business needs to know about the EU’s General Data Protection Regulation.
  \newblock Publications Office; 2018.
  
  \bibitem{lopez2006edugain}
  Lopez D, Solberg A, Stanica M.
  \newblock Edugain profiles and implementation guidelines.
  \newblock G{\'E}ANT2. 2006;.
  
  \bibitem{abac}
  Hu VC, Kuhn DR, Ferraiolo DF, Voas J.
  \newblock Attribute-Based Access Control.
  \newblock Computer. 2015;48(2):85--88.
  \newblock doi:{10.1109/MC.2015.33}.
  
  \bibitem{odrl}
  Ianella R.
  \newblock Open digital rights language (ODRL).
  \newblock Open Content Licensing: Cultivating the Creative Commons. 2007;.
  
  \bibitem{ipfs}
  Benet J. IPFS - Content Addressed, Versioned, P2P File System; 2014.
  \newblock Available from: \url{https://arxiv.org/abs/1407.3561}.
  
  \bibitem{go}
  Ashburner M, Ball CA, Blake JA, Botstein D, Butler H, Cherry JM, et~al.
  \newblock Gene Ontology: tool for the unification of biology.
  \newblock Nature Genetics. 2000;25(1):25--29.
  \newblock doi:{10.1038/75556}.
  
  \bibitem{do}
  Schriml LM, Arze C, Nadendla S, Chang YWW, Mazaitis M, Felix V, et~al.
  \newblock Disease Ontology: a backbone for disease semantic integration.
  \newblock Nucleic Acids Research. 2011;40(D1):D940--D946.
  \newblock doi:{10.1093/nar/gkr972}.
  
  \end{thebibliography}

%

\end{document}